\documentclass[reprint,superscriptaddress,showpacs,aps,aip]{revtex4-1}
\usepackage[pdftex]{graphicx}
\usepackage{amsmath}
\usepackage{amssymb}
\usepackage{placeins}
\usepackage{color}
\usepackage{xfrac}
\usepackage[letterpaper, total={7.28in, 9.55in}]{geometry}
\usepackage{hyperref}
\hypersetup{
    colorlinks=true,
    linkcolor=blue,
    filecolor=blue,      
    urlcolor=blue,
    citecolor=blue
}
\usepackage{pdfpages} 
\makeatletter 
\AtBeginDocument{\let\LS@rot\@undefined}
\makeatother
\usepackage{pgffor}
\newcommand{\etal}{\textit{et al.}}
\newcommand{\urusi}{URu$_2$Si$_2$}
\newcommand{\urusip}{URu$_2$Si$_{2-x}$P$_x$}

\begin{document}
\graphicspath{{.}{figures/}}
\hyphenation{a-na-lo-gy}
\vspace{5cm}
\title{~\\~\\~\\Collinear antiferromagnetic order in \urusip~ revealed by neutron diffraction}

\author{M. C. Rahn}
\email[]{marein.rahn@tu-dresden.de}
\affiliation{Los Alamos National Laboratory, Los Alamos, New Mexico 87545, USA}
\affiliation{Institute for Solid State and Materials Physics, Technical University of Dresden, 01062 Dresden, Germany}
\author{A. Gallagher}
\affiliation{National High Magnetic Field Laboratory, Florida State University, Tallahassee, Florida 32310, USA}
\author{F. Orlandi}
\affiliation{ISIS Facility, STFC, Rutherford Appleton Laboratory, Chilton, Didcot, Oxfordshire, OX11 0QX, United Kingdom\looseness=-1}
\author{D. D. Khalyavin}
\affiliation{ISIS Facility, STFC, Rutherford Appleton Laboratory, Chilton, Didcot, Oxfordshire, OX11 0QX, United Kingdom\looseness=-1}
\author{C. Hoffmann}
\affiliation{Oak Ridge National Laboratory, Oak Ridge, Tennessee 37830, USA}
\author{P. Manuel}
\affiliation{ISIS Facility, STFC, Rutherford Appleton Laboratory, Chilton, Didcot, Oxfordshire, OX11 0QX, United Kingdom\looseness=-1}
\author{R. Baumbach}
\affiliation{National High Magnetic Field Laboratory, Florida State University, Tallahassee, Florida 32310, USA}
\author{M. Janoschek}
\email[]{marc.janoschek@psi.ch}
\affiliation{Los Alamos National Laboratory, Los Alamos, New Mexico 87545, USA}
\affiliation{Laboratory for Neutron and Muon Instrumentation, Paul Scherrer Institute, CH-5232 Villigen, Switzerland\looseness=-1}
\affiliation{Physik-Institut, Universit\"{a}t Z\"{u}rich, CH-8057 Z\"{u}rich, Switzerland}

\date{\today}

\begin{abstract}
The hidden order phase in \urusi~is highly sensitive to electronic doping. A special interest in silicon-to-phosphorus substitution is due to the fact that it may allow one, in part, to isolate the effects of tuning the chemical potential from the complexity of the correlated $f$ and $d$ electronic states. We investigate the new antiferromagnetic phase that is induced in URu$_2$Si$_{2-x}$P$_x$ at $x\gtrsim0.27$. Time-of-flight neutron diffraction of a single crystal ($x=0.28$) reveals $c$-axis collinear $\mathbf{q}_\mathrm{m}=(\sfrac12,\sfrac12,\sfrac12)$ magnetic structure with localized magnetic moments ($\sim$2.1--2.6\,$\mu_\mathrm{B}$). This points to an unexpected analogy between the (Si,P) and (Ru,Rh) substitution series. Through further comparisons with other tuning studies of \urusi, we are able to delineate the mechanisms by which silicon-to-phosphorus substitution affects the system. In particular, both the localization of itinerant 5$f$ electrons as well as the choice of $\mathbf{q}_m$ appear to be consequences of the increase in chemical potential. Further, enhanced exchange interactions are induced by chemical pressure and lead to magnetic order, in which an increase in inter-layer spacing may play a special role.
\end{abstract}

\maketitle

\section{Introduction}
The complexity of energetically nearly degenerate electronic states in strongly correlated materials often gives rise to unusual ordering phenomena and exotic physics~\cite{Dagotto2005, Maple:10}. However, it is challenging to identify the hierarchy of the underlying interactions between charge, orbital, magnetic, and structual degrees of freedom. Such is the case for the hidden order (HO) state that emerges in the heavy fermion material \urusi. Studies using a large set of external parameters to tune this system have revealed a rich phase space of adjacent ordered phases~\cite{Mydosh2011, Mydosh2014,Mydosh2020}, many of which are magnetic. Notably, characteristics and symmetry of the HO state itself are known to be markedly different from conventional spin or charge orders in strongly correlated metals~\cite{Kung2015}. Even though the onset of HO at $T_0$ = 17.5\,K is marked by a second-order symmetry breaking phase transition~\cite{Palstra:85, Maple:86}, the true symmetry of the associated  order parameter remains elusive.

In \urusi, strong electronic correlations arise due to the hybridization of localized uranium $f$ electrons with the conduction electrons, as evident from a large single-ion Kondo temperature $T_K=120$\,K \cite{Schmidt:10, Aynajian:10}, and the formation of a coherent Kondo lattice at $T^\ast$~$\approx70$\,K \cite{Rodrigo:97, Palstra:86, Schoenes:87}. The onset of HO is accompanied by the opening of a charge gap over about 40\% of the Fermi surface (FS), as observed via various methods sensitive to changes in the band structure \cite{Palstra:85,Maple:86,Schoenes:87,Oh:07, Kasahara:07, Bonn:88, Ohkuni:99, Altarawneh:11}. This reorganization of the electronic structure below $T_0$ was originally attributed to the emergence of charge or spin density wave order, which would be typical of itinerant magnetism. More detailed investigations using modern electronic band structure methods revealed a secondary hybridization of a heavy $f$-like quasiparticle band with a light holelike band at $Q^\ast=\pm0.3\pi/a$, which results in the formation of a hybridization gap $\Delta_{Q^\ast}=5$\,meV~\cite{Santander-Syro:09, Schmidt:10, Aynajian:10}. Both angle-resolved photoemission spectroscopy (ARPES) \cite{Bareille:14} and neutron spectroscopy~\cite{Wiebe:07, Butch:15} even demonstrate that larger parts of the Fermi surface are gapped. Nuclear magnetic resonance (NMR) measurements \cite{Shirer:13} also indicate the presence of a pseudogap below 30 K.

This dramatic reorganization of the Fermi surface suggests that the HO order is a result of its intricate electronic band structure. However, the complex metallic state of \urusi~also exhibits strong anisotropy in the spin and charge channels, which is typically associated with localized electronic degrees of freedom\cite{Altarawneh:12}. Torque magnetization~\cite{Okazaki:11}, high-resolution x-ray diffraction~\cite{Tonegawa:14}, elastoresistance~\cite{Riggs:14}, NMR~\cite{Kambe:13}, and Raman spectroscopy~\cite{Kung2015} indicate that the electronic state breaks the tetragonal symmetry of the underlying crystal structure, which led to the proposal that the HO state may be of nematic origin. Even though other x-ray diffraction\cite{Tabata:14}, NMR~\cite{Kambe:2018}, and thermodynamic~\cite{Wang:2020} studies at ambient pressure have not corroborated this tetragonal symmetry breaking, a recent x-ray diffraction study revealed a tetragonal-to-orthorhombic phase transition as a function of pressure~\cite{Choi2018}. Furthermore, ultrasound measurements have observed an orthorhombic lattice instability due to a volume-conserving strain field with $\Gamma_3$ symmetry~\cite{Yanagisawa2013}. Taken together, this suggests that the difference between the various studies may be due to varying crystal quality, resulting in different amounts of internal strain.  

The above underscores that unraveling the conundrum of HO requires a better understanding of the underlying duality of the itinerant and localized degrees of freedom. To this end, it is helpful to examine the impressive collection of studies in which the HO state has been tuned by various control parameters. This includes the external parameters of high magnetic fields~\cite{Aoki2012,Levallois2009}, pressure and strain~\cite{Chandra2002,Amitsuka2007}, as well as chemical substitution, on the uranium ~\cite{Delatorre1992, Ocko1997} or ruthenium site~\cite{Dalichaouch:1989, Dalichaouch1990, Dalichaouch:1990, Bauer:05,Butch:09,Burlet1992, Kawarazaki1994, Kuwahara2013, Prokes2017a, Prokes2017b, Kanchanavatee2011, Kanchanavatee2014}.

In all of these cases, it is observed that HO exists in close proximity to magnetic phases, which, in many cases, resemble those found in other tetragonal members of the U$T_2$Si$_2$ ($T$: transition metal) family~\cite{Umarji1987,Mihalik2006,Verniere1996,Ning1992,Matsuda2003}. While this provides some insight, it is often unclear how to disentangle the effects of varying the hybridization (e.g. by varying the degree of delocalization and spin-orbit coupling of the ligand), of the local environment (by variation of bond lengths and angles) and the variation of the Fermi surface. This is exemplified by a recent magnetoresistivity study under combined high pressures and  magnetic fields that finds that the effects of both tuning parameters are intertwined~\cite{Knafo2020}. To address this issue, the (Si,P) substitution series has been established~\cite{Baumbach2014,Gallagher2016,Gallagher2016a,Shirer2017,Wartenbe2017}, which is thought to weaken $p$--$f$ hybridization~\cite{Chappell2020}, but affects the spacing and orientation of $d$-ligands only weakly. With the character of $f$-$d$ interaction held intact, the donation of one $p$ electron is thought to emphasize the effects of varying the chemical potential in the~\urusi~host. Even though the various consequences of any chemical substitution are necessarily intertwined, the special significance of the (Si,P) series is that it provides a different hierarchy in which the numerous relevant energy scales of \urusi~are affected. One may therefore hope that this will provide the necessary contrast to disentangle the mechanisms by which the HO state is manipulated in other tuning studies of this material.

\begin{figure}
\includegraphics[width=0.9\columnwidth,trim= 0pt 0pt 0 0pt, clip]{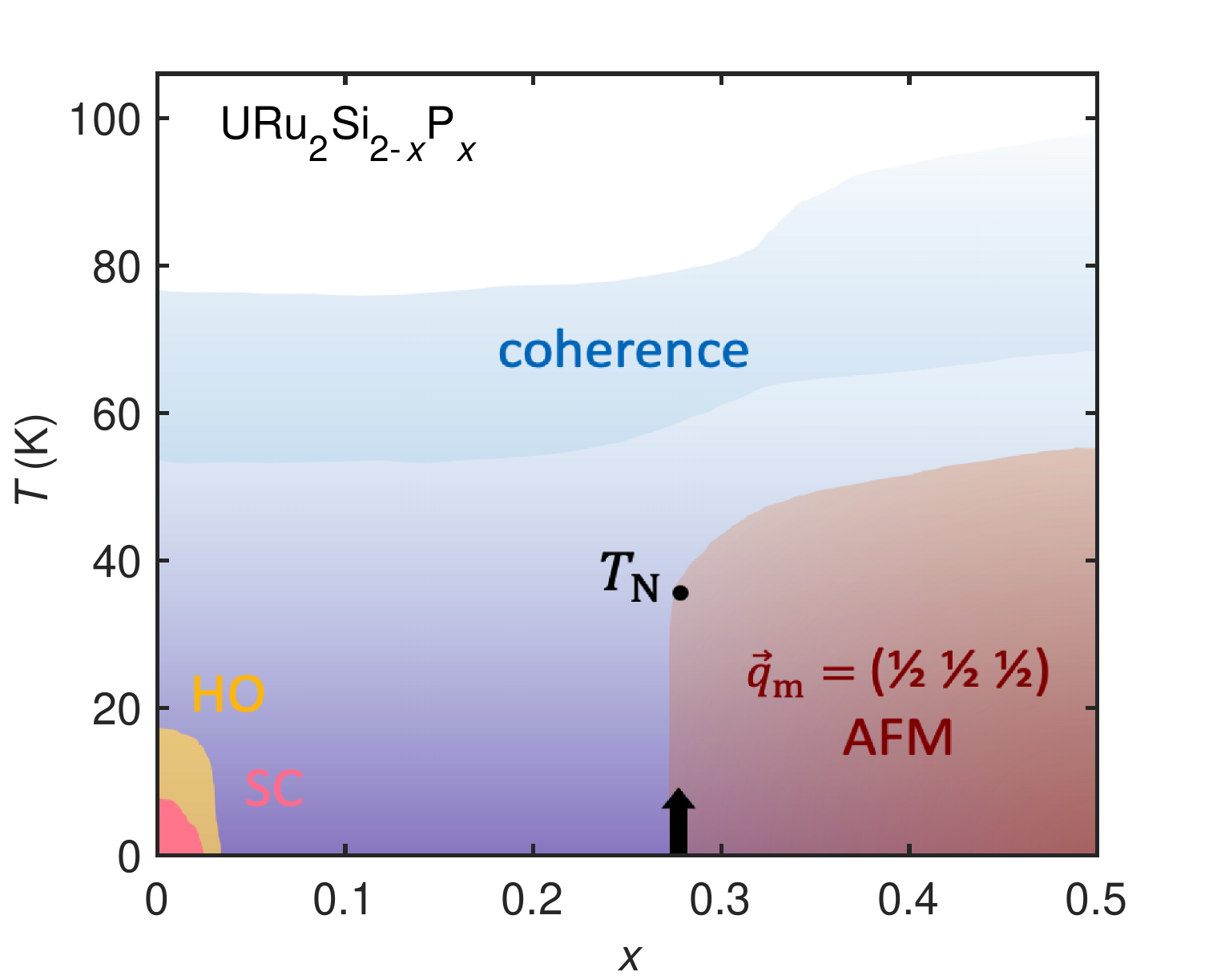}
\caption{\label{Fig1} Simplified phase diagram of \urusip, adapted from Gallagher~\etal, \cite{Gallagher2016}.  Few percents of phosphorus substitution suppress the hidden order (HO) and, with it, superconductivity (SC). The arrow marks the composition of the long range antiferromagnetically (AFM) ordered sample investigated in this study.}
\end{figure}

An overview of the effects of (Si,P) substitution is given by the schematic phase diagram in Fig.~\ref{Fig1}. Interestingly, superconductivity (SC) and HO prove to be highly sensitive to very small P doping levels. In particular, the superconducting critical temperature $T_c\approx1.4\,$K weakly increases to a maximum at $x\approx0.01$, before suppression of SC at $x\approx0.028$ and suppression of HO at $x\approx0.035$~\cite{Gallagher2016a}. Quantum oscillation measurements in this regime indicate that no significant changes of the Fermi surface are associated with the destruction of the HO phase~\cite{Huang2019}. Following a paramagnetic Kondo lattice state in the range $0.035\lesssim x \lesssim 0.26$, antiferromagnetism is abruptly stabilized at $x\gtrsim 0.27$~\cite{Gallagher2016}. To better understand the different roles of the large number of available tuning parameters of HO in~\urusi, it is of great interest to characterize the order in this new magnetic phase~\cite{Mydosh2020}. Here we report on a neutron diffraction study carried out to determine its antiferromagnetic order.

\begin{figure*}
\includegraphics[width=2.\columnwidth,trim= 0pt 0pt 0pt 0pt, clip]{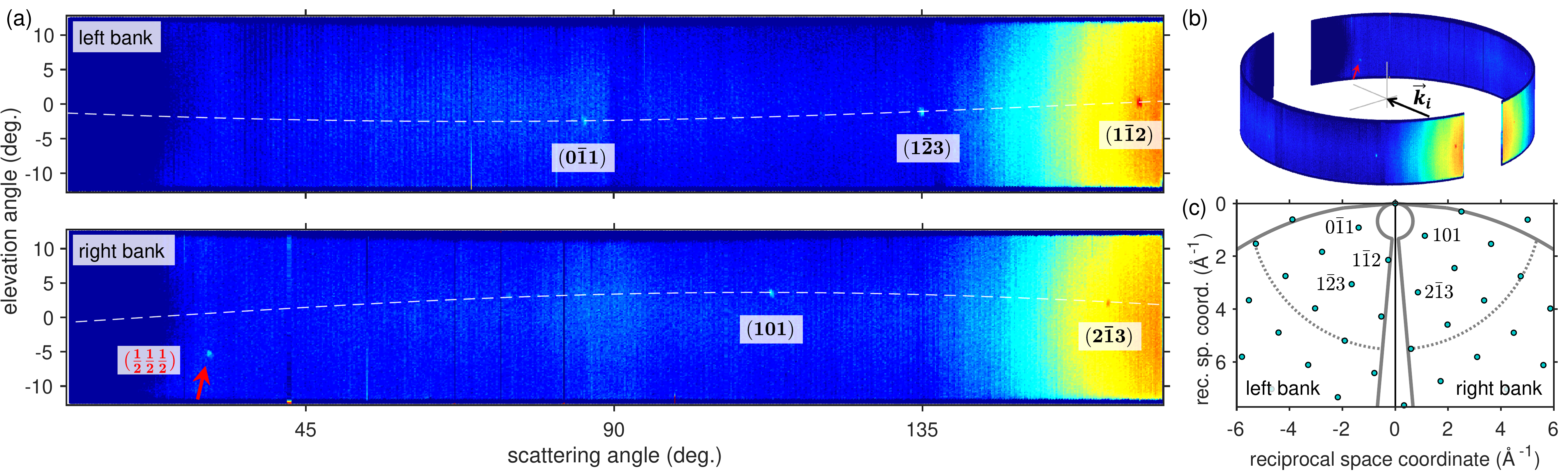} 
\caption{\label{Fig2} (a) Overview of Bragg peaks observed at WISH, at 2\,K. The upper (lower) panel shows the raw neutron counts detected on the left (right) detector bank, on an arbitrary logarithmic scale. The $(101)$--$(0\bar{1}1)$ plane of reciprocal space is indicated by a white dashed line. The $(1,1,1)$ direction, which features the only observed magnetic Bragg peak, is seen below this plane, at a scattering angle of around 30$^\circ$. (b) Perspective view of this data, illustrating the layout of the instrument. (c) Schematic view of the $(101)$--$(0\bar{1}1)$ plane of reciprocal space (Note that the magnetic peak at $\mathbf{q}_m$ is observed below this plane). The accessible range of momentum transfers is delineated by a broad gray line and peaks seen in (a) are labeled in analogy. The dotted line indicates the momentum transfer at which the magnetic form factor of uranium has decreased by $1/e$ ($\approx5.5$\,\AA$^{-1}$).}
\end{figure*}

\section{Experimental Details}
Single crystals of \urusip were synthesized by an indium flux method~\cite{Baumbach2014,Gallagher2016a}. Aside from the high purity of the resulting crystals, this method also overcomes the issue of the high vapor pressure of phosphorus. The phosphorus concentration $x=0.28(1)$ of this sample (i.e. $\approx$14\,\% substitution) was determined by energy-dispersive x-ray spectroscopy (EDX). The uncertainty of this value was estimated by performing a number of measurements on different positions of the sample surface. For reference, this composition is marked by a black arrow in the phase diagram in Fig.~\ref{Fig1}. The magnetic susceptibility $\chi(T)$ of this crystal was measured using a superconducting quantum interference device (SQUID) magnetometer (Quantum Design), in a field of 0.5\,T applied either parallel or perpendicular to the $c$ axis. The dimensions of single crystals grown by the molten metal flux technique make magnetic neutron diffraction barely feasible. We selected a crystallite for its large size compared to the average sample yield, with a mass of only $\approx0.5$\,mg and dimensions of $0.8\times0.8\times0.05$\,mm$^3$. 

The issue of small sample size can be overcome using the latest generation time-of-flight neutron diffractometers, which combine high-brilliance neutron moderators with highly optimized focusing neutron guides. The high flux yield at the sample position enables experiments on single crystals with dimensions of $\sim$1\,mm and less. In combination with detector banks that collect scattered neutrons over a large solid angle, this allows for experiments that were impossible until recently. Preliminary measurements of the nuclear scattering down to 90\,K were carried out at the TOPAZ instrument at SNS (Oak Ridge National Laboratory), which receives neutrons from a decoupled poisoned hydrogen moderator. The investigation of the magnetic order at 2\,K was performed at the WISH instrument at the ISIS pulsed neutron source (STFC, Rutherford Appleton Laboratory)~\cite{Chapon2011}. WISH looks onto a solid methane (40\,K) moderator, which provides high-brilliance neutron pulses with a broad band of wavelengths from 1 to 10~\AA. Neutrons are collected on a detector bank that continuously covers a wide range of scattering angles ($10^\circ\leq 2\theta \leq 170^\circ$) with 1\,m tall position-sensitive $^3$He detectors. WISH also employs an oscillating radial collimator that defines a cylindrical collimated area in the center of the sample tank, which provides the low background required for studies with such small samples. The crystal was mounted in a dedicated low background cryostat (Oxford Instruments) with the [110] and [101] directions in the scattering plane. The broad margin of accessible out-of-plane momentum transfers covered a volume equivalent to (more than) one full Brillouin zone, which was crucial for the identification of the magnetic propagation vector. An illustration of the accessible range of reciprocal space resulting from this configuration is given in Fig.~\ref{Fig2}. 

\section{Results}
\begin{figure}
\includegraphics[width=1\columnwidth,trim= 0pt 0pt 0 0pt, clip]{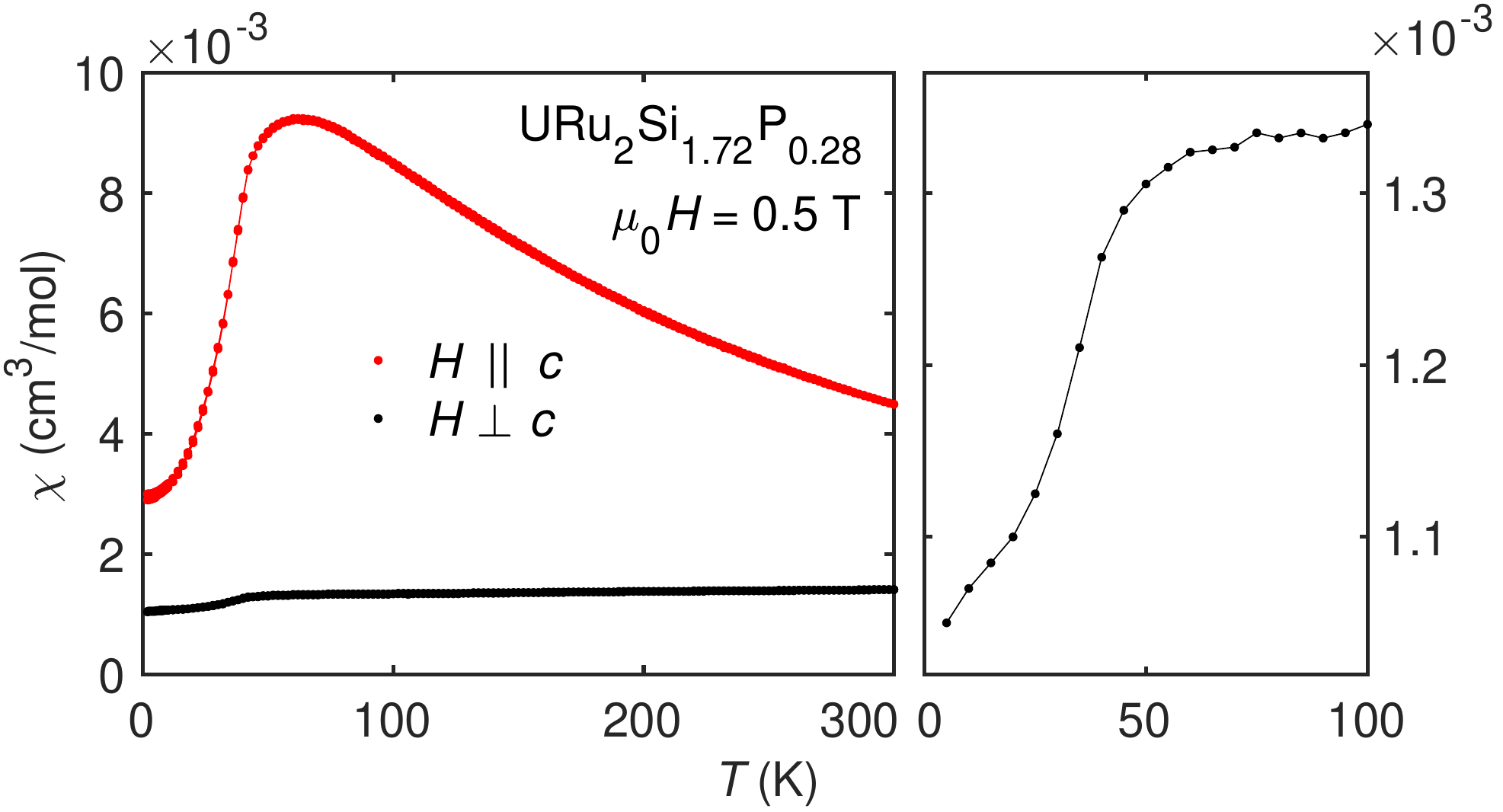}
\caption{\label{chi} Magnetic susceptibility of the \urusip~($x=0.28$)  crystal investigated by neutron diffraction, in a field of 0.5\,T applied either parallel or perpendicular to the $c$ axis. The right panel shows a detailed view of the $\mathbf{H}\perp\mathbf{c}$ data. As in the parent compound, the characteristics are dominated by the onset of Kondo screening around $T_\mathrm{coh}\approx80$\,K, as well as a strong $c$-axis single ion anisotropy.}
\end{figure}

Figure~\ref{chi} shows the temperature dependence of the magnetic susceptibility $\chi(T)$ in a magnetic field of 0.5\,T, applied parallel and perpendicular to the $c$ axis. The characteristics are dominated by the strong single ion magnetic anisotropy, which indicates the same $c$-axis Ising character known of the parent compound~\cite{Palstra:85}. The broad maximum around  $T_\mathrm{coh}\approx 80$\,K marks the onset of Kondo screening of the magnetic moments. The magnetic phase transition is associated with a marked decrease in $\vec{H}\perp \vec{c}$ susceptibility. These observations are also consistent with recent NMR measurements, which revealed a commensurate internal field $H_\mathrm{int}\approx0.85$\,kOe oriented along the $c$ direction in the antiferromagnetic state of \urusip~\cite{Shirer2017}. 

With the sample orientation and time-of-flight range of the WISH experiment illustrated in Fig~\ref{Fig2}, the accessible nuclear Bragg peaks were indexed in the $I4/mmm$ unit cell of the parent compound (lattice parameters $a=4.12$\,\AA~and $c=9.57$\,\AA). The scale factor of the nuclear intensities was refined with the Rietveld method using FullProf~\cite{Rodriguez1993} after a single-crystal Lorentz correction had been performed in Mantid~\cite{Arnold2014}. Variables of this fit included the vertical position of Si/P ions [$z=0.38(1)$] and a parameter controlling the extinction correction~\cite{Rodriguez1993}. The (Si,P) stoichiometry was fixed to the value determined by EDX [$x=0.28(1)$]. A comparison of measured and calculated intensities is shown in Fig.~\ref{Fig4}. Numerical values and a detailed account of this fit is provided in the Supplemental Material~\cite{SM}.

\begin{figure}
\includegraphics[width=0.95\columnwidth,trim= 0pt 0pt 0 0pt, clip]{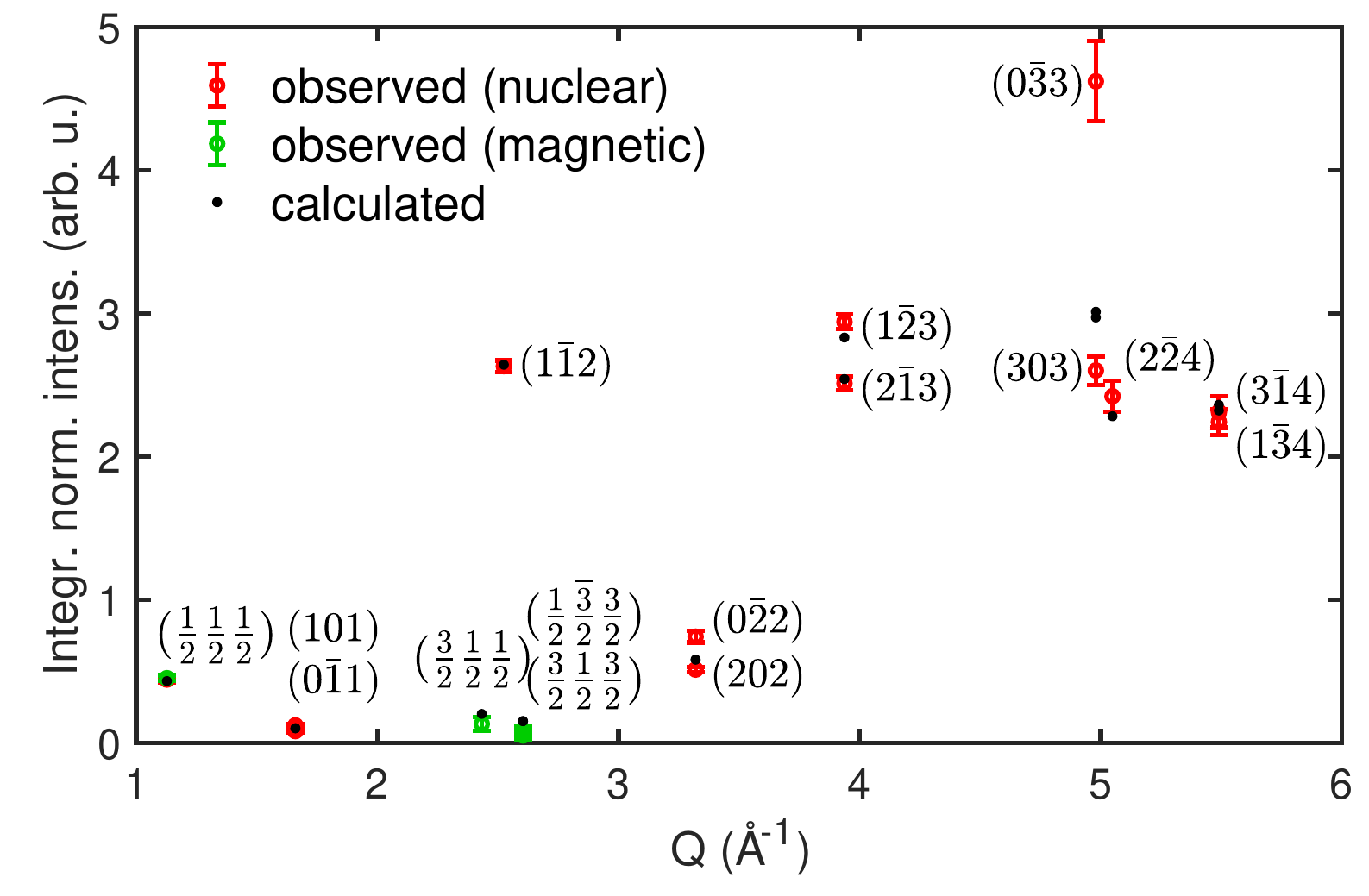}
\caption{\label{Fig4}Comparison of Bragg intensities calculated for \urusip~ ($x=0.28$) with those measured at WISH at 2\,K. The overall scale factor was inferred from a refinement of nuclear intensities (red). The magnitude of the ordered magnetic moment was then fitted separately to reproduce the intensities of magnetic reflections (green). Numerical values of these fits are given in the Supplemental Material~\cite{SM}.}
\end{figure}

The intensity of these reflections was tracked between 2 and 80\,K. At low temperatures, our measurements reveal a magnetic Bragg peak at momentum transfer $\mathbf{Q}=(\sfrac12,\sfrac12,\sfrac12)$, which corresponds to the magnetic propagation vector $\mathbf{q}_\mathrm{m}=\mathbf{Q}$. The widths of this peak in reciprocal space were the same as for nuclear peaks, which indicates that the range of the order is not limited by the coherence length of magnetic correlations. Other instances of scattering from the $\mathbf{q}_\mathrm{m}$ vector were also identified in higher-order Brillouin zones, although these intensities are increasingly suppressed by the magnetic form factor.

\begin{figure}
\includegraphics[width=0.95\columnwidth,trim= 0pt 0pt 0 0pt, clip]{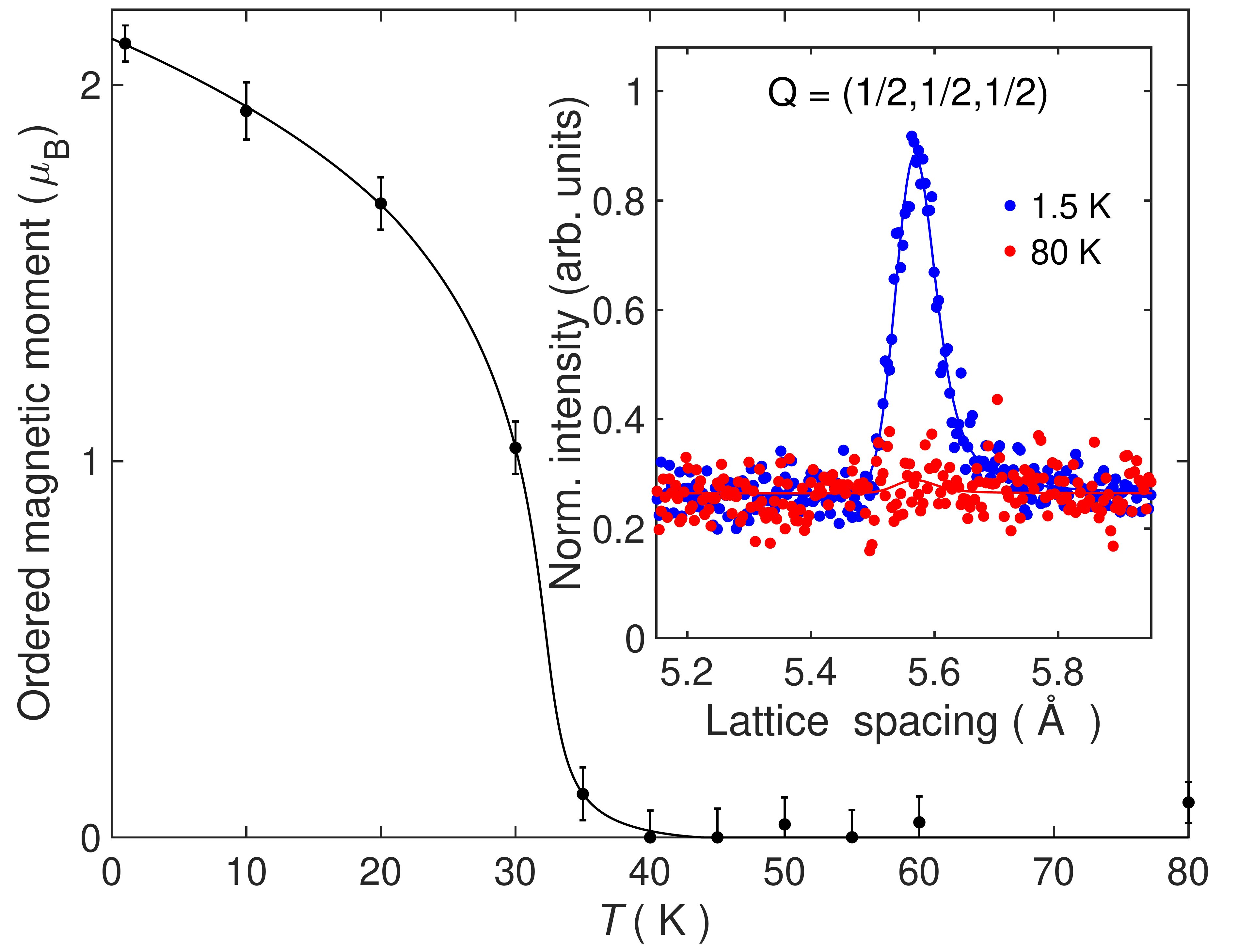}
\caption{\label{Fig5} Temperature dependence of the ordered magnetic moment in~\urusip~($x=0.28$), with an ordering temperature of $T_\mathrm{N}=32.6(7)$\,K and critical exponent $\beta=0.31(4)$. The inset illustrates the emergence of a magnetic Bragg reflection at momentum transfer $\mathbf{Q}=\mathbf{q}_\mathrm{m}=(\sfrac12,\sfrac12,\sfrac12)$.}
\end{figure}

Figure~\ref{Fig5} shows the temperature dependence of the integrated intensity of this peak, on a scale of the estimated ordered magnetic moment $M$ per uranium ion (see below). A fit of the temperature-dependence of the ordered moment via $M(T)\propto(1-\frac{T}{T_\mathrm{N}})^{\beta}$ yields a N\'{e}el temperature of $T_\mathrm{N}=32.5(1.3)\,$K and a critical exponent of $\beta=0.24(6)$. The large uncertainty notwithstanding, this is in line with the three-dimensional (3D) Ising character ($\beta_{\textrm{th.}}=0.32$)~\cite{chaikin1995} and the strong magnetic anisotropy evident from Fig.~\ref{chi}.

\begin{figure}
\includegraphics[width=0.7\columnwidth,trim= 185pt 155pt 1130pt 80pt, clip]{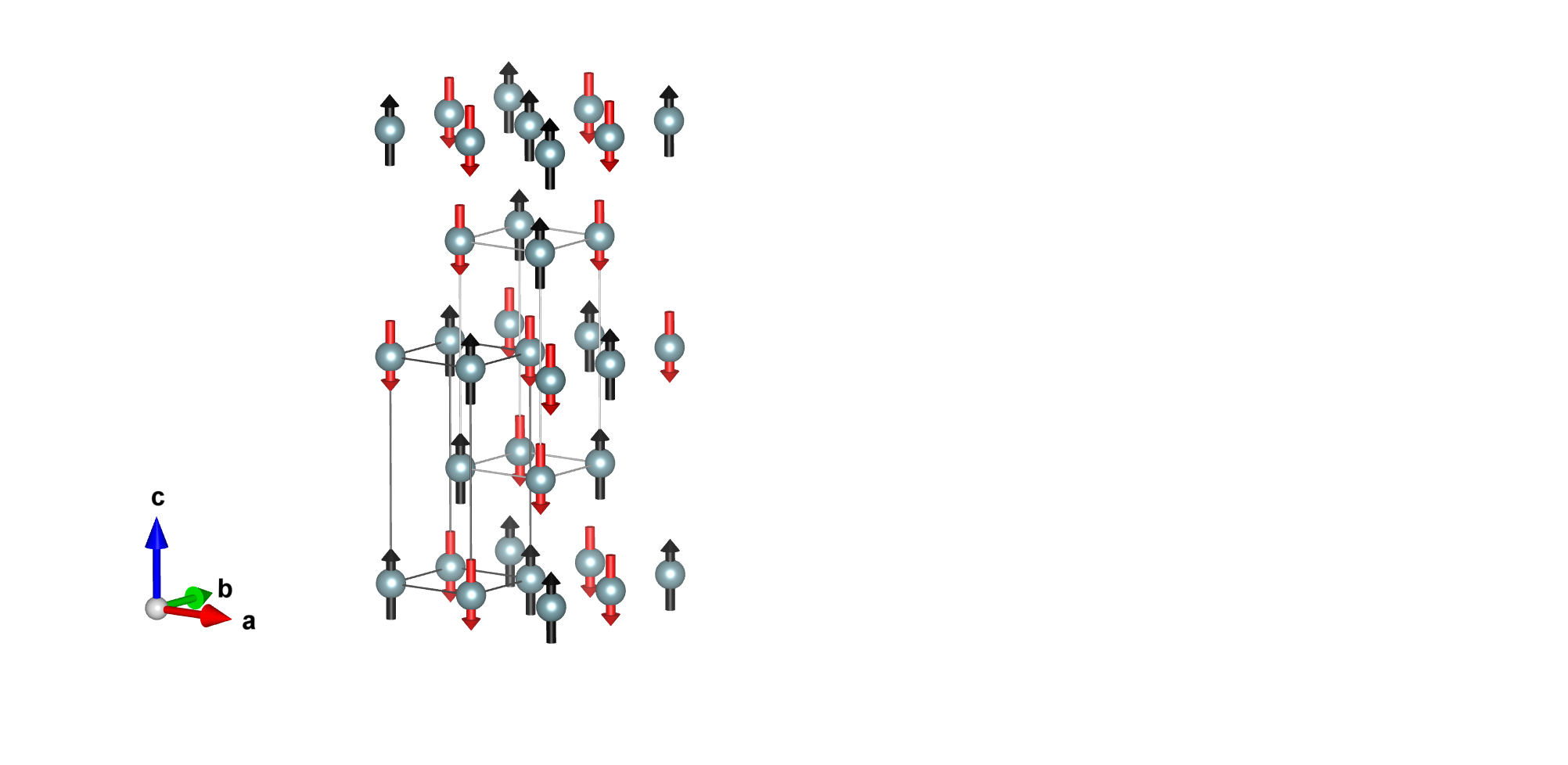}
\caption{\label{Fig6} $c$-axis collinear $\mathbf{q}_\mathrm{m}=(\sfrac12,\sfrac12,\sfrac12)$ magnetic structure of ~\urusip~($x=0.28$), described by the Shubnikov group $I_c4_1/acd$ (No. 142.570). For clarity, only uranium ions are shown and opposite spins are drawn in different colors.}
\end{figure}

\section{Analysis and interpretation}
The present study represents a favorable case in which the magnetic structure is fully determined by the observation of a single magnetic peak, given the constraints inferred from magnetometry and symmetry. Representational analysis was performed using the ISODISTORT program~\cite{Campbell2006,Stokes2019}. The propagation vector $\mathbf{q}_m=(\sfrac12,\sfrac12,\sfrac12)$ represents the $P$ point (``k12'') in the Brillouin zone of space group $I4/mmm$. For magnetic moments at the uranium site, this yields two magnetic irreducible representations (irreps), mk12t2 and mk12t5 (Kovalev notation), with magnetic moments along the $c$ axis and in plane, respectively.

Each irrep provides a choice of order parameter directions, corresponding to magnetic space groups. The resulting magnetic structures are illustrated in Fig.~S1 of the Supplemental Material~\cite{SM}. Since magnetometry clearly indicates $c$-axis Ising anisotropy (cf. Fig.~\ref{chi}), any magnetic space group corresponding to mk12t5 (for which all moments are confined to the $a$--$b$ plane) can be ruled out. Of the three choices for mk12t2 (P1, P3 and C1), P3 is not physical as it forces one U site to be non-magnetic, and C1 represents the unlikely case that the size of the ordered magnetic moment varies between the two U ions (of the same Wyckoff site, i.e. in an environment of the same point symmetry). There is then indeed only one possible solution, mk12t2 P1, which corresponds to the magnetic space group $I_c4_1/acd$ (No. 142.570).

The resulting magnetic structure is illustrated in Fig.~\ref{Fig6}. The basis and origin of the $I_c4_1/acd$ magnetic cell, specified in terms of lattice vectors of the parent paramagnetic space group $I4/mmm$, are $(\bar{1},1,0)$, $(\bar{1},\bar{1},0)$, $(0,0,2)$, and $(\sfrac14,\sfrac14,\bar{\sfrac74})$, respectively. It is the $\bar{4}m'2'$ symmetry of the Wyckoff site $8a$ which constrains the magnetic dipole moments to align with the $c$ axis.

By comparing the integrated magnetic and nuclear intensities, we obtain an ordered magnetic moment of $\mu\sim 2.1$--$2.6\,\mu_\mathrm{B}$. This range reflects the systematic uncertainty of the intensity scale factor (cf. Fig.~\ref{Fig4} and Ref.~\onlinecite{SM}). By comparison, the uncertainty due to the  choice of the neutron magnetic form factor~\cite{Kuwahara2006} is not significant (U$^{3+}$ vs. U$^{4+}$ , d$\mu\approx0.04\,\mu_\mathrm{B}$). This result is consistent with local-moment-like antiferromagnetic phases in related U$A_2B_2$ compounds~\cite{Ptasiewicz1981,Durakiewicz2014}. 

\section{Discussion}
Our neutron diffraction study demonstrates that the magnetically ordered state observed in \urusip~ for phosphorus concentrations  $x\gtrsim 0.27$~is described by a $c$ axis collinear antiferromagnetic structure with propagation vector $\mathbf{q}_\mathrm{m}=(\sfrac12,\sfrac12,\sfrac12)$ and an ordered moment of $\sim$2.1--$2.6\,\mu_\mathrm{B}$. The commensurate character of this state suggests that it arises from local uranium magnetic moments. This is corroborated by a recent NMR study of an $x=0.33$ single crystal that evidenced a homogeneous antiferromagnetic state emerging due to the localization of $5f$-electrons at higher P concentrations~\cite{Shirer2017}, as well as measurements of the Sommerfeld coefficient of the specific heat, which is reduced sharply once magnetic order emerges for $x\gtrsim 0.27$\cite{Gallagher2016}. In this section, we highlight the most relevant similarities and differences between our findings and other tuning studies of \urusi~\cite{Mydosh2020}.

Most importantly, the magnetic state discovered in the (Si,P) system reveals an unexpected parallel to the phase diagram of URu$_{2-x}$Rh$_x$Si$_2$~\cite{Burlet1992, Kawarazaki1994}. In this series, the HO vanishes for $x \gtrsim 0.08$. As in the (Si,P) system, this is followed by a paramagnetic heavy Fermi liquid region, for $0.08\lesssim x \lesssim 0.18$~\cite{Burlet1992, Kawarazaki1994,Prokes2017a}. Finally, for $0.18\lesssim x \lesssim 0.7$, a similar long-range $\mathbf{q}_\mathrm{m}=(\sfrac12,\sfrac12,\sfrac12)$ (or $\mathbf{q}_\mathrm{m}=(\sfrac12,\sfrac12,L)$~\cite{Kawarazaki1994}) antiferromagnetic state appears, with $\mu\approx2\,\mu_\mathrm{B}$ and $T_\mathrm{N}$ up to 44\,K, as in the present case.

An equivalence of (Si,P) and (Ru,Rh) substitution is far from obvious. For example, one may expect that the main effect of substituting U or Ru (e.g., by Np, Fe, Os, Ru, Rh) may be to alter the $d$--$f$ hybridization. It was only pointed out very recently that the case of (Si,P) may actually have similar consequences, given that the radial contraction of the $p$-orbitals weakens the $p$--$f$ hybridization~\cite{Chappell2020}.

On the other hand, both the (Si,P) and (Ru,Rh) series are markedly set apart from isoelectronic chemical substitutions (i.e., by Fe, Os, and Ge, respectively). The latter have a stronger impact on bond lengths and angles, as well as on spin-orbit coupling (SOC)~\cite{Wolowiec2020}. By comparison, structural modifications in the (Si,P) series are more gentle and the increase in SOC is negligible~\cite{Gallagher2016}. On the other hand, a considerable consequence of adding $p$ electrons to the system must be the rise in chemical potential.

Taken together, this suggests that (Si,P) and (Ru,Rh) substitution affect the HO state of \urusi~in two ways: by moderate chemical pressure and electron donation. It is then interesting to trace the roles of these two effects in stabilizing $\vec{q}_m=(\sfrac12,\sfrac12,\sfrac12)$ large moment magnetic order.

On the one hand, the increase of the chemical potential binds the $f$-electron states well below the Fermi energy, favoring their localization. On the other hand, at higher $x$, the decrease in unit cell volume becomes more relevant, as it increases the exchange integrals between neighboring $5f$ orbitals, promoting long range magnetic order. This increase of correlations at higher substitution levels also enhances the coherence temperature, as observed in electrical transport and magnetic susceptibility measurements (discussed below in the context of Fig.~\ref{Tcoh})\cite{Gallagher2016}. 

\begin{figure}
\includegraphics[width=0.9\columnwidth,trim= 0pt 0pt 0pt 0pt, clip]{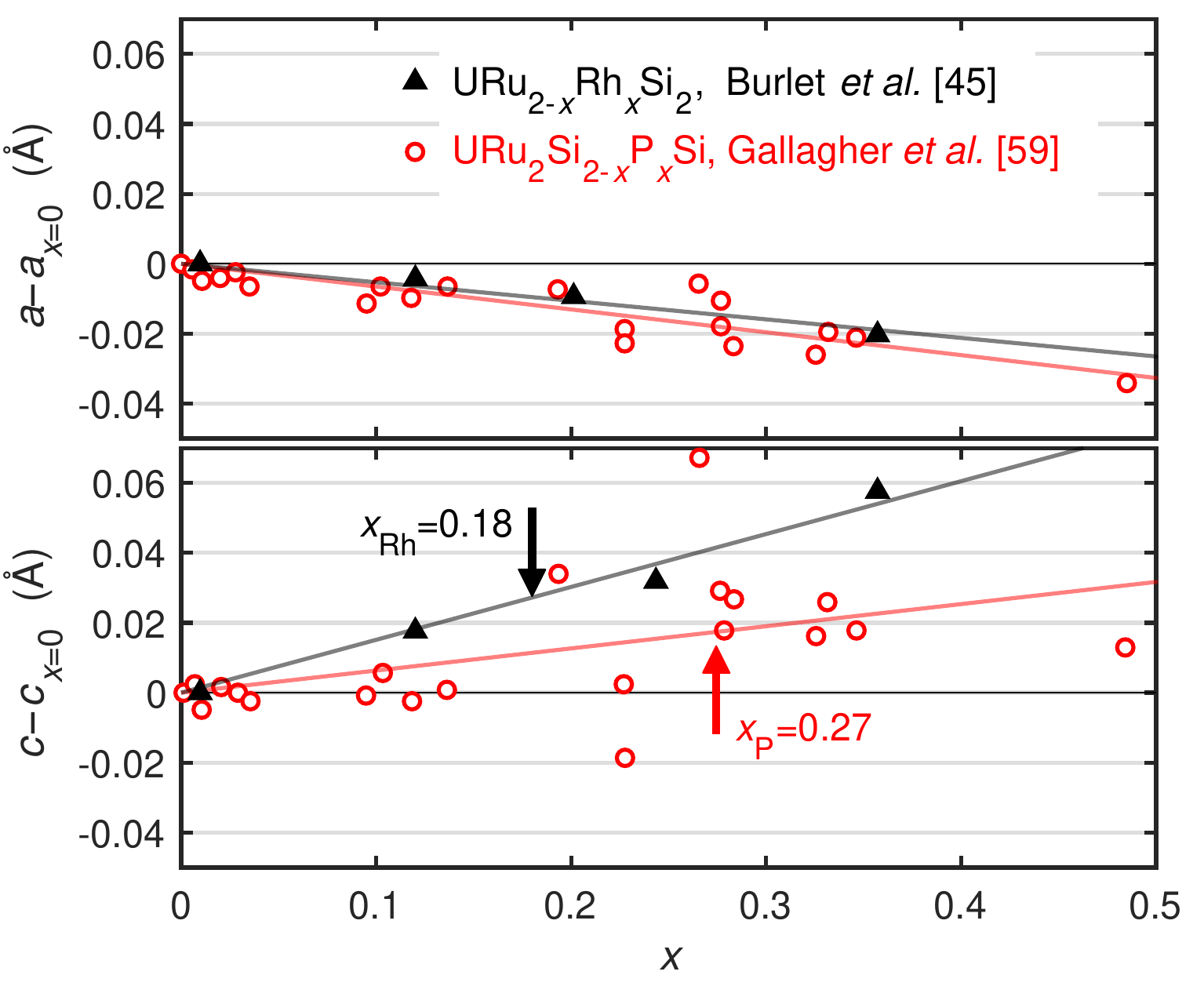}
\caption{\label{dadc} Changes of the lattice parameters $a$ (top) and $c$ (bottom) in the (Ru,Rh) and (Si,P) substitution series. The data are adapted from studies by Burlet~\etal, Ref.~\cite{Burlet1992} and Gallagher~\etal, Ref.~\cite{Gallagher2016}. In the lower panel, the critical compositions at which local moment $\mathbf{q}_m=(\sfrac12,\sfrac12,\sfrac12)$ magnetic order emerges in either series are marked by arrows.}
\end{figure}

The amount of chemical pressure exerted by (Si,P) and (Ru,Rh) substitution is similar. In Fig.~\ref{dadc}, we compare the variation of the lattice parameters in the two doping series, based on data reported by Burlet~\etal~\cite{Burlet1992} and Gallagher~\etal~\cite{Gallagher2016}. The contraction of the basal plane (parameter $a$, corresponding to the nearest-neighbor spacing of uranium ions) is in fact identical within the uncertainties of the measurements. On the other hand, the $c$-axis expansion (which likely acts as a handle on interlayer correlations) due to (Ru,Rh) substitution is significantly larger than in the (Si,P) series. Notably, the values of $c$ at which long-range order sets in are similar in both compounds (cf. arrows in Fig.~\ref{dadc}). Even though the number of outliers in the $c-c_{x=0}$ data by Gallagher~\etal~calls for caution, this makes for a tentative explanation for the difference in the corresponding critical substitution levels. 

In this context, studies of \urusi~under applied hydrostatic pressure are of special interest because they allow one to single out the effects attributable to structural variations. Applied pressure drives a quantum phase transition from the HO state to an antiferromagnetic phase with propagation vector $\mathbf{q}_\mathrm{m}=(0,0,1)$ at a critical pressure $P_\mathrm{c}=$ 0.7--1.0\,GPa~\cite{Amitsuka:99} (see, also, Ref.~\onlinecite{Butch:10} and references therein). At ambient pressure, neutron scattering originally observed a similar magnetic phase, however, with a much reduced magnetic moment of about 0.01\,$\mu_\mathrm{B}$~\cite{Broholm:87}. One important conclusion of the extensive efforts to investigate \urusi~under pressure is that this small moment antiferromagnetism (SMAF) at ambient pressure is likely a parasitic effect, induced locally by remnant strain~\cite{Niklowitz:10}. The closing of a spin gap at $\mathbf{q}_\mathrm{m}$, as observed in neutron scattering~\cite{Villaume:08}, emphasizes that the HO state is fundamentally different from the large moment pressure-induced antiferromagnetic phase. It is also consistent with the view that pressure enhances exchange interactions.

Chemical and applied pressure can be compared quantitatively using the Birch-Murnaghan equation of state [$P_\mathrm{chem}\approx9\,\textrm{MPa}\,(\Delta V/V)$], as reported by Gallagher~\etal~\cite{Gallagher2016}. In Fig.~\ref{Tcoh}, we use this relation to compare the variation of $T_\mathrm{coh}$ in the (Si,P) series (inferred from the broad maximum in $\chi(T)$, data adapted from Ref.~\onlinecite{Gallagher2016}) to the corresponding results of a high-pressure \urusi~study by Pfleiderer~\etal~\cite{Pfleiderer2006} (the same effect is also observed in resistivity measurements~\cite{Kagayama1994}). The two means of compressing the lattice indeed increase the coherence temperature by similar amounts. This suggests that even though the resulting antiferromagnetic structures differ, similar physics is at play in stabilizing the local moment order parameter. This is also supported by the observation that for the P concentration at which the antiferromagnetic phase sets in, the lattice contraction corresponds to a chemical pressure of $\approx0.75$\,GPa~\cite{Gallagher2016}, which is comparable to $P_\mathrm{c}$~\cite{Amitsuka:99,Butch:10}.

To recapitulate, (Si,P) and (Ru,Rh) substitution affect the lattice in a similar way, and both induce a $(\sfrac12,\sfrac12,\sfrac12)$ large moment magnetic order. The equivalent amount of applied pressure similarly induces a large moment antiferromagnetic state in \urusi, albeit with propagation vector $(0,0,1)$. The fact that the donation of one electron is the common difference between these substitution series and applied pressure suggests that the variation of the chemical potential has the role of selecting the different magnetic symmetry.

In the Supplemental Material~\cite{SM}, we also provide a discussion of the effect of applied magnetic fields (with reference to the studies [\onlinecite{Scheerer2012,Kim:03,Harrison2013,Scheerer2014,Ran2017,Das2015}]). This is another tuning parameter that may serve to reveal similarities and differences between different ground states of \urusi~derivatives. Here, we merely note that, as in \urusip, an antiferromagnetic state can be induced in the paramagnetic regime of URu$_{2-x}$Rh$_x$Si$_2$ by applying a critical field of 26\,T. A pulsed-field neutron diffraction study has shown that this magnetic order corresponds to a commensurate up-up-down ferrimagnetic structure with propagation vector $\mathbf{q}_\mathrm{IN}=(\sfrac13,0,0)$ \cite{Kuwahara2013, Prokes2017b}. Since the high-field magnetization of P and Rh substituted samples in the paramagnetic regime have very similar characteristics~\cite{Wartenbe2017}, and the magnetic order at higher P and Rh concentrations is identical, we speculate that the field-induced magnetic phase in \urusip~is also the one described by $\mathbf{q}_\mathrm{IN}$.

In summary, it is not straightforward to identify the mechanism of the large moment antiferromagnetic state discovered in \urusip~at $x\gtrsim0.27$, since the consequences of ionic substitution are necessarily entangled. However, such a large catalog of tuning studies is now available~\cite{Mydosh2020} that it becomes possible to recognize the key effects by comparison. In the present case, we thus arrive at a simplified picture in which the rise in chemical potential forms local magnetic moments (associated with the destruction of HO), before correlations are increasingly enhanced by chemical pressure, eventually stabilizing long-range magnetic order. The chemical potential is likely decisive in selecting a propagation vector that is distinct from the large moment phase of pure \urusi~under pressure.

\begin{figure}
\includegraphics[width=0.9\columnwidth,trim= 0pt 0pt 0pt 0pt, clip]{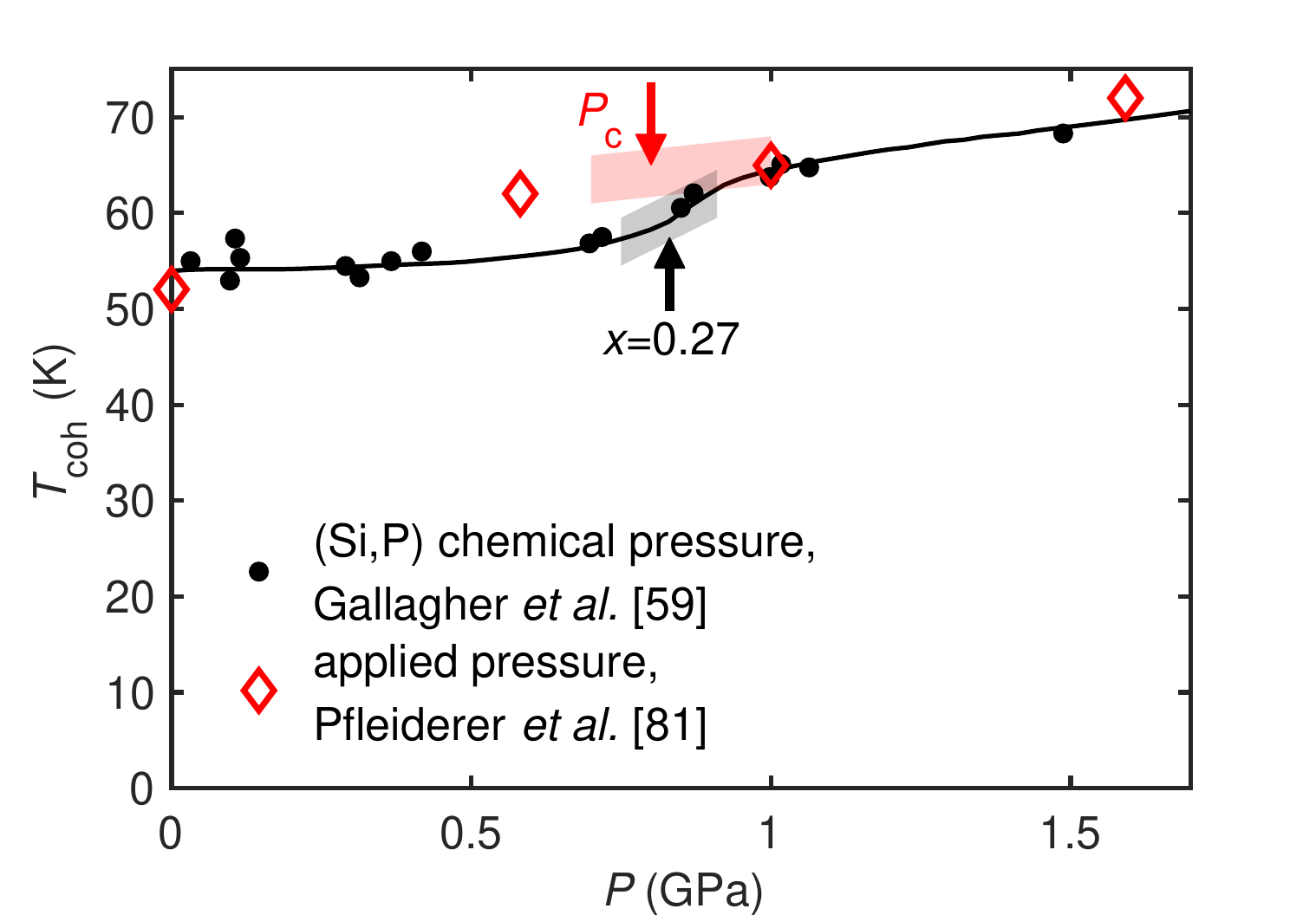}
\caption{\label{Tcoh} Pressure dependence of the heavy-fermion coherence scale $T_\mathrm{coh}$, inferred from a broad maximum in magnetic susceptibility $\chi(T)$ curves (cf. Fig.~\ref{chi}). Data for chemical pressure in the present doping series, adapted from Ref.~\cite{Gallagher2016}, is compared to measurements of the parent compound under applied pressure, reported by Pfleiderer~\cite{Pfleiderer2006}. The arrows and shaded margins indicate the regimes where long range magnetic order is induced by hydrostatic (red) and chemical (black) pressure.}
\end{figure}

\section{Conclusion}
Using state-of-the-art neutron time-of-flight diffractometers, we were able to determine the magnetic structure in a minute single crystal of \urusip~ ($x>0.27$). Our measurements indicate $c$-axis collinear antiferromagnetic order with a propagation vector $\mathbf{q}_\mathrm{m}=(\sfrac12,\sfrac12,\sfrac12)$ and an ordered moment of $\sim$2.1--$2.6\,\mu_\mathrm{B}$. This highlights that the phase diagrams of the two substitution series \urusip~\cite{Gallagher2016,Gallagher2016a,Shirer2017,Wartenbe2017} and URu$_{2-x}$Rh$_x$Si$_2$~\cite{Burlet1992, Kawarazaki1994, Kuwahara2013, Prokes2017a, Prokes2017b} are nearly identical with respect to the observed sequence of ground states. By comparison of various tuning studies, we infer that the localization of 5$f$ electrons as well as the selection of the $\mathbf{q}_\mathrm{m}=(\sfrac12,\sfrac12,\sfrac12)$ order parameter is a common consequence of the increased chemical potential, whereas enhanced exchange interactions are induced by chemical pressure, in which the increase in interlayer spacing may play a special role.

In spite of these parallels, we note that the (Si,P) and (Ru,Rh) substitutions must act differently in terms of the spin-orbit coupling, lattice strain, and local crystal electric field, and may alter different aspects of the Fermi surface. Detailed investigations of the underlying electronic structures and local degrees of freedom would be in place to shed more light on these differences.

It is also important to keep in mind that the observed behavior near quantum phase transitions between different correlated phases is known to be highly sensitive to the ionic disorder introduced by chemical substitution. It has been shown to impact both electrical transport properties~\cite{Rosch1999} as well as the nature of the quantum phase transition itself~\cite{Kirkpatrick2015, Huang2013, Huang2016}. In turn, the impact of disorder remains another important open question when comparing these two substitution series.

Neutron diffraction or NMR measurements of \urusip~with $x \gtrsim 0.26$ as a function of magnetic field would be of great interest to confirm whether the high-field induced magnetic order in this system~\cite{Wartenbe2017} is indeed the same as that found in the (Ru,Rh) series~\cite{Prokes2017b}. 

Finally, it is worth highlighting that extensive work on \urusi~has demonstrated that the various forms of magnetic order that emerge by destabilizing the HO phase are representative of magnetism in the extended family of U$T_2$Si$_2$ compounds ($T$: transition metal). This poses the fascinating question of whether HO may be stabilized in these related materials as well.
~\\
\begin{acknowledgments}
Work performed by RB and AG at the National High Magnetic Field Laboratory was supported by National Science Foundation Cooperative Agreement No. DMR-1644779 and the State of Florida. The synthesis of crystalline materials was supported by the Center for Actinide Scienceand Technology (CAST), an Energy Frontier Research Center (EFRC) funded by the U.S. Department of Energy (DOE), Office of Science, Basic Energy Sciences (BES), under AwardNo. DE-SC0016568. Work by M.J. at Los Alamos National Laboratory was supported by the U.S. Department of Energy, Office of Basic Energy Sciences, Division of Materials Science and Engineering, under project ``Quantum Fluctuations in Narrow-Band Systems.'' Part of this research was carried out at the ISIS Neutron Facility, an institution of the UK Science and Technology Research Council (STFC). Research conducted at SNS (TOPAZ instrument) was sponsored by the Scientific User Facilities Division, Office of Basic Energy Sciences, U.S. Department of Energy. M.C.R. is grateful for fellowships provided by the LANL Director's Fund and the Alexander von Humboldt Foundation. Work by MCR at TU Dresden was supported by the Deutsche Forschungsgemeinschaft through the CRC 1143 and the Würzburg-Dresden Cluster of Excellence ct.qmat (EXC 2147, Project ID 390858490).
\end{acknowledgments}

\bibliography{URu2SiP2}



\section*{Supplementary material}

# Supplemental Material:

# Colinear antiferromagnetic order in URu$_2$Si$_{2-x}$P$_x$ revealed by neutron diffraction

M. C. Rahn, A. Gallagher, F. Orlandi, D. D. Khalyavin, C. Hoffmann, P. Manuel, R. Baumbach, and M. Janoschek

**Appendix A: Extended Discussion — URu$_2$Si$_{2-x}$P$_x$ in magnetic fields**

The application of strong magnetic fields along the $c$-axis of URu$_2$Si$_2$ yields in a more complex phase diagram, with a cascade of first-order phase transitions at fields $H_1 = 35$ T, $H_2 = 36/37$ T (up/down ramp), and $H_3 = 39$ T. $H_1$ marks the suppression of the HO, and $H_3$ denotes the transition to a field-polarized paramagnetic regime [1, 2]. A recent neutron diffraction study in pulsed fields demonstrated that the three field-induced magnetic phases are described by spin density wave (SDW) order with $\mathbf{q}_{\mathrm{SDW}} = (0.6, 0, 0)$, which has been associated with the putative nesting vector of the Fermi surface. A series of quantum oscillation measurements has shown that the transitions $H_1$ to $H_3$ are marked by reconstructions of the Fermi surface [3–6].

In the present context, it is interesting to discuss how this high-magnetic field phase diagram is altered by phosphorus substitution. A recent electrical transport and magnetization study in magnetic fields up to 65 T [7] finds that for $x \lesssim 0.035$ the highest-field SDW phase becomes more pronounced as a function of $x$. For the regime $0.035 \lesssim x \lesssim 0.26$, where no ordered state exists at low fields, only a single field-induced phase is observed, albeit over a substantially broader field-range of $30 \sim 45$ T. Finally, for $x \gtrsim 0.26$, the local moment antiferromagnetic order that we have identified here appears to persist up to about 45 T. It is followed by a field-induced phase with similar magnetotransport features as for the SDW order identified for lower $x$, and extends up to about 60 T. We note that, by contrast, when URu$_2$Si$_2$ is tuned with high magnetic fields and applied pressure, the field-induced phases are relatively quickly suppressed as a function of pressure [8].

Due to similarities in high-magnetic field magnetization data, Wartenbe *et al.*[7] have compared the antiferromagnetic phase of URu$_2$Si$_{2-x}$P$_x$ investigated here with the isoelectronic (Ru,Fe) substitution [9]. However, as our work shows, the magnetic phases underlying the observed behavior in magnetic field are quite different. Notably, the magnetic phase stabilized by Fe substitution is the same as found via application of pressure. In turn, it was suggested initially that the effects of Fe substitution are dominated by the resulting chemical pressure [10, 11]. This notion is complicated by the observation that isoelectronic substitution with Os, which should nominally act as negative chemical pressure, yields a similar phase diagram [12]. Instead of pressure, it may be that the $c/a$ lattice parameter ratio, which behaves similar for both series, controls this phase transition [12]. It has also been suggested that the large $5d$ orbital radius of Os may enhance the orbital overlap and hybridization in a way that outweighs the effects of lattice expansion [13].

**Appendix B: Refinement of integrated neutron scattering intensities**

To extract integrated neutron scattering Bragg intensities, the raw a data were normalized (to current and monitor) and a single crystal time-of-flight Lorentz correction was applied. For each peak, a region of interest was then defined on the detector surface and the signal was diffraction-focused onto one-dimensional datasets (i.e., intensity vs. wavelength or $d$-spacing). The intensities and standard deviations are then obtained by fitting a skew Lorentzian lineshape. We confirmed that this procedure was reliable and reproducible. The resulting integrated squared structure factors $F^2$ are summarized in the Table below, along with the corresponding calculated values.

Among the recorded nuclear Bragg peaks, we observed three instances in which nominally equivalent peaks appeared with disparate intensities. In our experiment, the relevance of self-absorption can be ruled out due to the tiny size of the crystal, and no direction-dependent absorption by the sample environment (e.g. if a cryomagnet had been in place) is expected.

| | | | $d\,(\text{Å})$ | $\lambda\,(\text{Å})$ | $F^2_{\text{obs}}$ | $F^2_{\text{calc}}$ |
|---|---|---|---|---|---|---|
| $\frac{1}{2}$ | $\frac{1}{2}$ | $\frac{1}{2}$ | 5.576 | 2.97 | 0.44(2) | 0.43 (*) |
| $\frac{3}{2}$ | $\frac{1}{2}$ | $\frac{1}{2}$ | 2.58 | 2.85 | 0.13(5) | 0.20 (*) |
| $\frac{1}{2}$ | $\frac{3}{2}$ | $\frac{3}{2}$ | 2.41 | 3.12 | 0.08(3) | 0.15 (*) |
| $\frac{1}{2}$ | $\frac{3}{2}$ | $\frac{3}{2}$ | 2.41 | 4.08 | 0.05(3) | 0.15 (*) |
| 1 | 0 | 1 | 3.785 | 6.28 | 0.08(1) | 0.0959 |
| 0 | $\bar{1}$ | 1 | 3.785 | 5.13 | 0.12(1) | 0.0968 |
| 1 | $\bar{1}$ | 2 | 2.489 | 4.93 | 2.63(4) | 2.6357 |
| 2 | 0 | 2 | 1.893 | 3.15 | 0.51(2) | 0.5777 |
| 0 | $\bar{2}$ | 2 | 1.893 | 2.56 | 0.74(4) | 0.5817 |
| 2 | $\bar{1}$ | 3 | 1.596 | 3.15 | 2.51(5) | 2.5403 |
| 1 | $\bar{2}$ | 3 | 1.596 | 2.94 | 2.94(5) | 2.8272 |
| 3 | 0 | 3 | 1.262 | 2.10 | 2.60(10) | 2.9746 |
| 0 | $\bar{3}$ | 3 | 1.262 | 1.71 | 4.62(28) | 3.0066 |
| 2 | $\bar{2}$ | 4 | 1.245 | 2.47 | 2.42(11) | 2.2835 |
| 3 | $\bar{1}$ | 4 | 1.145 | 2.19 | 2.24(9) | 2.3197 |
| 1 | $\bar{3}$ | 4 | 1.145 | 1.98 | 2.31(11) | 2.3581 |

**URu$_2$Si$_{2-x}$P$_x$**
$I4/mmm$ (#139), $Z = 2$
$a = 4.121$ Å, $c = 9.571$ Å
$z_{\text{Si,P}} = 0.364(3)$
$R_w(F^2) = 8.03$ cf. Ref. [14]

| ion | Wyck. | x | y | z | occ. (%) |
|---|---|---|---|---|---|
| U | 2a | 0 | 0 | 0 | 100 |
| Ru | 4d | 0 | $\frac{1}{2}$ | $\frac{1}{4}$ | 100 |
| Si | 4e | 0 | 0 | z | 86 |
| P | 4e | 0 | 0 | z | 14 |

TABLE I. (left) Results of the refinement of nuclear intensities observed at WISH. (right) List of calculated and observed structure factors, as shown in Fig. 4 of the manuscript. First, the nuclear reflections were refined to determine scale and extinction parameters. These were then held fixed in separate fits of the (much weaker) magnetic intensities (*). The large uncertainty of the scale parameter yields a wide range of the ordered magnetic moment ($2.1 \sim 2.6\,\mu_{\text{B}}$). The standard deviation of fitting the four magnetic peaks (ca. $\pm 0.1\,\mu_{\text{B}}$) is less relevant.

The issue is therefore most likely attributable to energy-dependent extinction. Due to extinction, the intensities of equivalent reflections measured at different scattering angles can differ in orders of magnitude, which is a known issue at WISH (due to its wide range of wavelengths, 0.35–10 Å.). Even though we a wavelength-dependent extinction correction was taken into account in our refinement (FullProf program, Ref. 14), this model may not give good results if only few reflections are available.

Notably, the purpose of our *refinement* is merely to infer the scale parameter. We then use this parameter to estimate the magnitude of the ordered magnetic moment. If anomalously large peaks are excluded from the refinement (which significantly improves the residual), the scale parameter decreases and the corresponding magnitude of the moment increases, from $2.1\,\mu_{\text{B}}$ up to $2.6\,\mu_{\text{B}}$. This can be considered a systematic uncertainty, as stated in the manuscript.

### Appendix C: Identification of the propagation vector

To search for magnetic intensity, we calculated the difference signal between data taken at base temperature and that at 40 K. For each pixel, we then summed the difference signal over the relevant range of TOF. This reveals that the peak at $\vec{Q} = \vec{q}_m = (1/2, 1/2, 1/2)$ is by far the dominant magnetic signal. We therefore optimized the orientation of the sample to provide good signal-to-noise at $\vec{q}_m$.

We have also confirmed that no additional intensity appears at integer Bragg positions. The detector and time-of-flight (TOF) range indeed cover the full Brillouin zone, although not all detector areas offer the same signal-to-noise. As it varies throughout reciprocal space, it is difficult to make a simple statement on the overall accuracy of this zero-measurement.

We also searched the 2 K dataset for other $\vec{q}_m$ magnetic peaks and indeed found faint magnetic signals in higher-order Brillouin zones, e.g. at (1.5,0.5,1.5) and (0.5,-1.5,1.5). At these positions, the signal is already strongly suppressed by the magnetic form factor and the peaks also lie at detector positions with poor signal-to-noise. Since these observations are indeed barely above the noise level, they would not aid the determination of the ordered magnetic moment.

**Appendix D: Magnetic symmetry analysis**

Single crystal neutron diffraction is a powerful method for magnetic structure determination. The present study is a favorable case in which the observation of a single peak, along with magnetic susceptibility measurements and symmetry analysis, allows the solution of the structure.

As noted in the manuscript, representational analysis was performed using the ISODISTORT program [15, 16]. The propagation vector $\mathbf{q}_m = (1/2, 1/2, 1/2)$ represents the $P$ point ("k12") in the Brillouin zone of space group $I4/mmm$. For magnetic moments at the uranium site, this yields two magnetic irreducible representations (irreps), mk12t2 and mk12t5 (Kovalev notation), with magnetic moments along the $c$-axis, and in-plane, respectively.

Each irrep provides a choice of order parameter directions, corresponding to magnetic space groups. The resulting magnetic structures are illustrated in Fig. S1. Here we have superposed arrows indicating possible magnetic basis vectors (distortion modes) in separate colors, see caption.

This shows that any magnetic space group corresponding to mk12t5 can be ruled out, given the $c$-axis Ising character inferred from magnetometry (cf. manuscript). Of the three choices for mk12t2 (P1, P3 and C1), P3 is not physical as it forces one U site to be non-magnetic, and C1 represents the unlikely case that the size of the ordered magnetic moment varies between the two U ions (of the same Wyckoff site, i.e. in an environment of the same point symmetry).

This leaves mk12t2 P1 ($I_c4_1/acd$, #142.570) as the only choice. Fig. S1 illustrates that it may be viewed as a special case of C1, with the same moment size at both sites. Notably, the relative orientation of spins on the $I4/mmm$ cell and those on the body-centering sublattice can still be viewed as arbitrary, as it is equivalent to a translation of the whole crystal by $(1/2, 1/2, 1/2)$. If $m_1$ and $m_2$ are the magnetic moments on either sublattice, the neutron scattering intensity of any magnetic Bragg peak is proportional to $(m_1^2 + m_2^2)$, which also reflects this point.

---

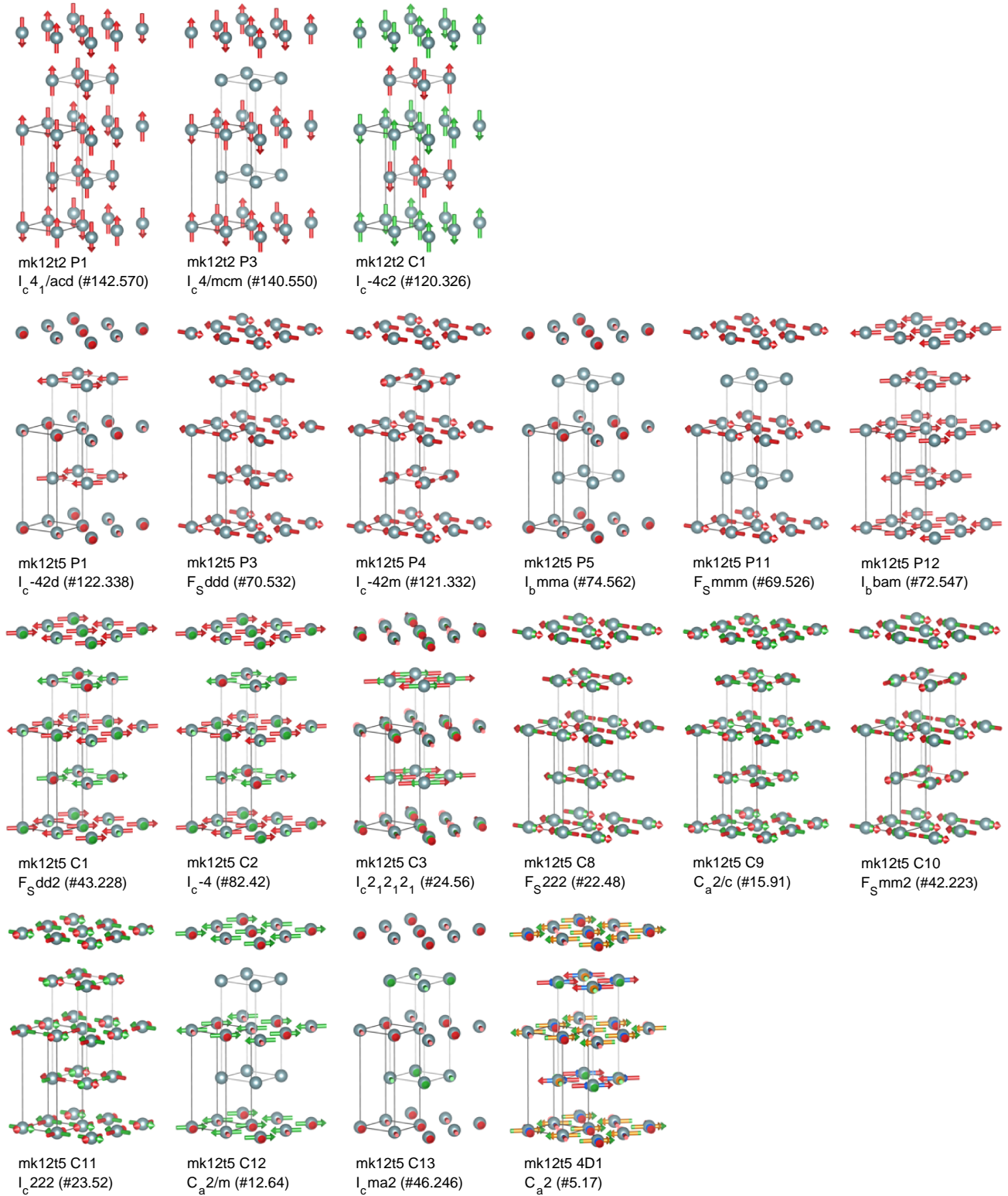

Fig. S1 Possible magnetic space groups inferred from representational analysis of $\mathbf{q}_m = (1/2, 1/2, 1/2)$, i.e. the $P$-point, or "k12" in space group $I4/mmm$. The three structures drawn in the top row correspond to the irrep mk12t2 (with magnetic moments constrained along the $c$-axis), and the remaining choices to irrep mk12t5 (with in-plane moments only). The magnetic distortion modes are indicated by superposed sets of colored vectors (there may be one or two, and, only in the case of mk12t5 4D1, four).

4



\end{document}